\documentclass[a4paper,aps,prl,groupedaddress,showkeys,showpacs, 
  superscriptaddress,floats,floatats,floatfix]{revtex4}
\usepackage[latin1]{inputenc}
\usepackage{graphicx}  
\usepackage{dcolumn}   
\usepackage{bm}        
\usepackage{natbib}
\usepackage{latexsym}
\usepackage{mathrsfs}
\usepackage{amssymb}
\usepackage{amsmath}
\usepackage{amscd}
\usepackage{color}
\usepackage{verbatim}
\begin{document}
\title{Dissipation effects in the ratchetlike Fermi acceleration} 
\author{Cesar Manchein and Marcus W.~Beims}
\email[E-mail address:~]{mbeims@fisica.ufpr.br}
\affiliation{Departamento de F\'\i sica, Universidade Federal do Paran\'a,
         81531-980 Curitiba, PR, Brazil}
\date{\today}

\begin{abstract}
Ac driven asymmetric pulses can be used to control the Fermi acceleration
between three different motions, ${\bf A:}$ the {\it accelerated} mode;
${\bf D:}$ the {\it decelerated} mode and ${\bf H:}$ the {\it hyperaccelerated} 
mode. In this work we show that dissipation strongly affects the particles 
velocity, reducing the possibility for an accurate control of the dynamics. 
The saturation time, where the mean velocity starts 
to be constant due to dissipation, decays with a power law
$\sim\gamma^{-\beta}$, where $\gamma$ is the dissipation parameter and
$\beta$ is close to $1$. The value of the saturated mean velocity also 
decays with a power law with exponent $\beta\sim0.6$ for the case ${\bf H}$, 
and  $\beta\sim0.3$ for the case ${\bf A}$. In the case  ${\bf D}$ this
velocity is almost constant for small dissipations.
\end{abstract}



\maketitle

Fermi acceleration is a topic which got attention in various areas of physics, 
ranging from nonlinear physics 
\cite{lichtenberg72,lichtenberg92,leonel04,leonel04-prl,leonel05,egydio06,
karlis06,karlis07}, atom optics \cite{steane95,schleich98,saif05}, plasma 
physics \cite{milovanov01,michalek99} to astrophysics 
\cite{veltri04,kobayakawa02,malkov98}. After the first model proposed 
by Fermi~\cite{fermi49}, essentially two different versions became common
in the literature. In the first one, the Fermi-Ulam (FU) model, a bouncing 
particle moves between a fixed surface and a parallel oscillating surface
\cite{ulam61}. In this case the regular islands in the phase space prevent 
the Fermi acceleration.
A simplified version of this model was proposed to improve simulations 
\cite{lichtenberg72}, called the static wall model. It essentially ignores 
the displacement of the moving wall but keeps the information for 
the momentum transfer as the wall was oscillating. 
The dynamics of the static model was studied in different aspects 
\cite{lichtenberg72,leonel04,leonel04-prl,leonel05} and for different 
models~\cite{egydio06}, and the relevant result for the purpose of the
present work is that invariant curves in the phase 
space, found for higher velocities, prevent the particle to increase its 
kinetic energy without bounds. Recently the hopping wall 
approximation was proposed \cite{karlis06,karlis07} which takes into
account the effect of the wall displacement, and allows the analytical 
estimation of the particle mean velocity. Compared to the simplified
static model, the particle acceleration is enhanced. The second kind of 
Fermi accelerated model was proposed in 1977 by Pustyl'nikov 
\cite{pustylnikov77}, who considered a particle on a periodically 
oscillating horizontal surface in the presence of a gravitational field. 
Different from the FU model, for some initial conditions and control 
parameters the particle energy can grow indefinitely. 

In this paper we analyze the effect of dissipation in the simplified FU 
model when an ac driven asymmetric pulse controls the Fermi acceleration 
(deceleration) \cite{cesarJCP09}. Small dissipation is inevitable in real 
systems and its influence on the dynamics of conservative systems is of 
great interest \cite{grebogi96,rech05,janePD} since elliptic periodic 
orbits become small sinks and attractors start to exists 
\cite{feudel96,ricardo}. 
In the context of Fermi models dissipation effects have been inserted in 
two ways: frictional force \cite{leonel08} and inelastic 
collisions at the walls \cite{egydio07,leonelBJP08,leonelCHAOS07}. Here we
use the second approach and the 
pulse is a {\it deformed} sawtooth driving law for the moving wall. 
This Ratchetlike pulse differs from the ac driven asymmetric pulses 
({\it symmetric} 
sawtooth) used for the Fermi acceleration in the early work of Lichtenberg 
{\it et.~al} \cite{lichtenberg72} and proposed recently to control the motion of 
magnetic flux quanta \cite{cole06} and to analyse the relative  efficiency of 
mechanism leading to increased acceleration in the hopping wall approximation 
\cite{karlis07}. In the simplified Fermi model~\cite{lichtenberg72} the
particle is free to move between the elastic impacts with the walls. Consider 
that the moving wall oscillates between two extrema with amplitude $v_0$.
The gravitational force is considered zero. The system is described by a 
two-dimensional map 
$M_{1(2)}(V_n,\phi_n)=(V_{n+1},\phi_{n+1})$ which gives, respectively, the 
velocity of the particle, and the phase of the moving wall, immediately after 
the particle suffers a collision with the wall.  Considering dimensionless
variables the dissipative FU map with the {\it deformed} sawtooth wall is 
written as 

\begin{equation}
M_1:\left\{
\begin{array}{ll}
  V_{n+1} = \left\vert (1-\gamma)V_{n}+\dfrac{v_0}{\eta_1}(\phi_n-\eta_1) 
  \right\vert, \\
  \phi_{n+1} = \phi_{n} + \mu\dfrac{(\eta_1+\eta_2)}{V_{n+1}} 
\hspace{7mm} \mbox{mod}~(\eta_1+\eta_2),  
\end{array}
\right. 
\label{map_1}
\end{equation}
for $\phi_n<\eta_1$, and
\begin{equation}
M_2:\left\{
\begin{array}{ll}
  V_{n+1} = \left\vert (1-\gamma)V_{n}+\dfrac{v_0}{\eta_2}(\phi_n-\eta_1) 
  \right\vert, \\
  \phi_{n+1} = \phi_{n} + \mu\dfrac{(\eta_1+\eta_2)}{V_{n+1}} \hspace{7mm}
  \mbox{mod}~(\eta_1+\eta_2), 
\end{array}
\right. 
\label{map_2}
\end{equation}
for $\phi_n\ge\eta_1$, 
where $n$ is the iteration number and $\mu$ is the maximum distance between the 
walls. For $\gamma\ne 0$ we have dissipation effects.
Since in this simplified model the displacement of the moving wall is
ignored, the modulus function is used to avoid errors due to successive
collisions which may occur in the original model. In other words, if after 
a collision with the wall the particle continues to have a negative velocity
(a successive collision will occur in the original model), the particle moves
beyond the wall. The modulus for the velocity injects the particle back and
fixes the problem.

\begin{figure}[htb]
\begin{center}
 \includegraphics*[width=6.0cm,angle=0]{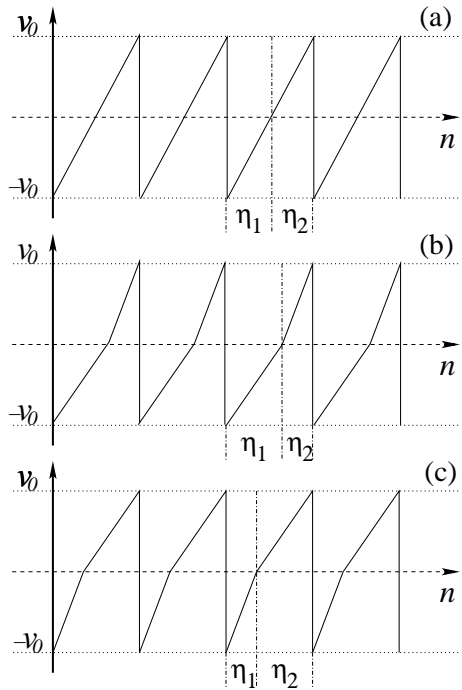}
\end{center}
\caption{\sl (a)-(c) The shape of the pulses used in the simulations. The 
deformed sawtooth effect is obtained when $\Delta\eta=\eta_2-\eta_1\ne0.0$. 
For (a) $\Delta\eta=0$ we have the symmetric {\it accelerated} case ${\bf A}$,
(b) $\Delta\eta<0$ we have the deformed {\it decelerated} case 
${\bf D}$ and (c) $\Delta\eta>0$ the deformed {\it hyperaccelerated} case 
${\bf H}$.}
\label{bilhar}
\end{figure}

\begin{figure}[htb]
\begin{center}
 \includegraphics*[width=8.0cm,angle=0]{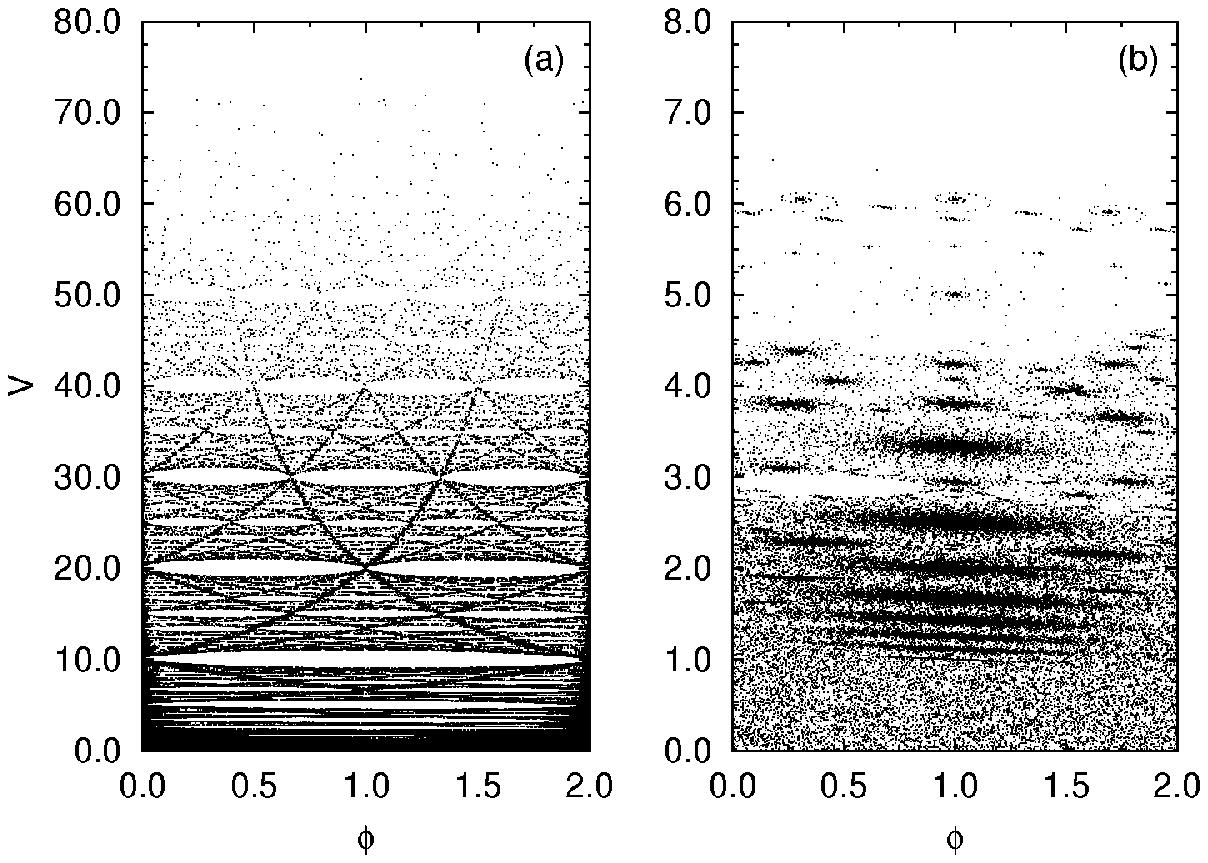}
 \includegraphics*[width=8.0cm,angle=0]{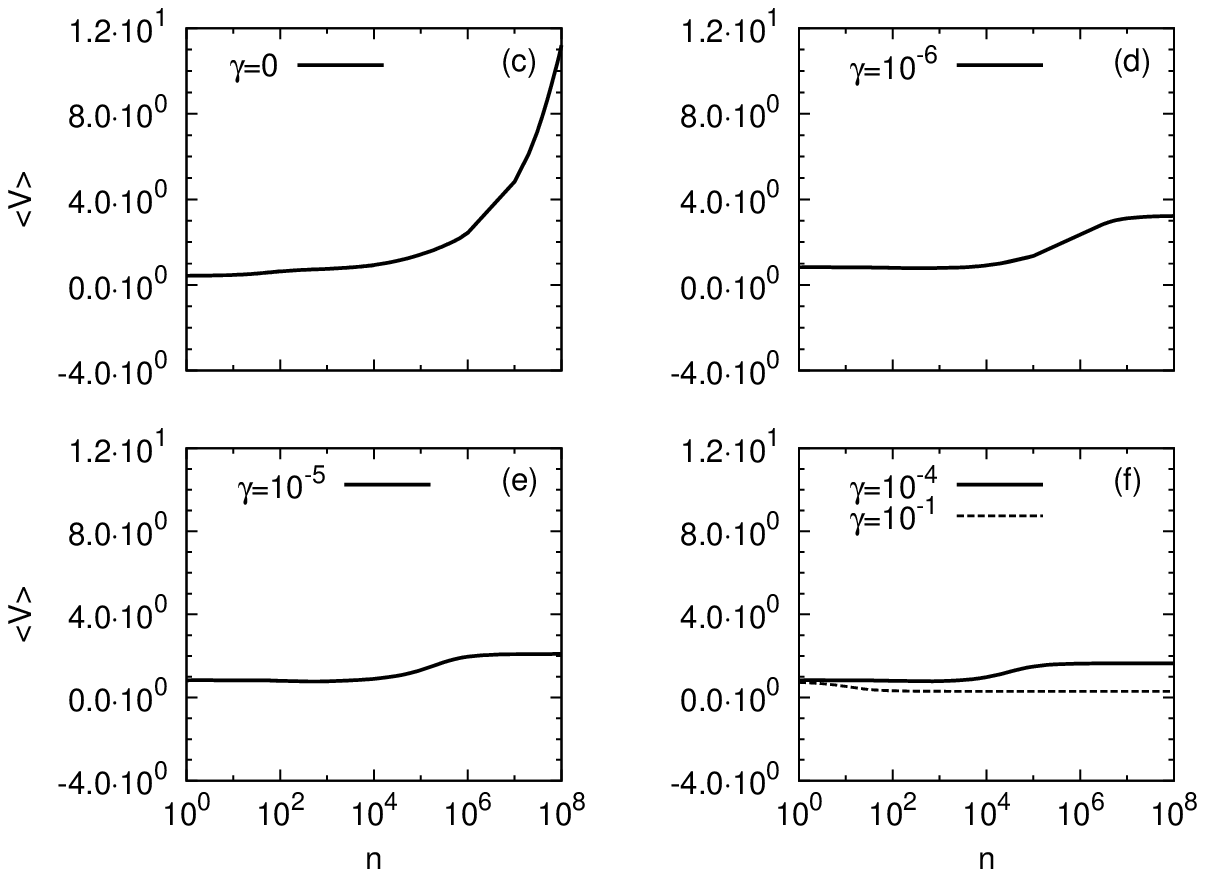}
\end{center}
\caption{\sl Evolution of $500$ chaotic orbits on the phase space
 $V\times\phi$ for the parameters $\mu=10$, $v_0=0.2$ for case
${\bf A}\, (\Delta\eta=0.0,\, \eta_1=\eta_2=1.00$). For Fig.~(a) 
$\gamma=0.0$ and (b) 
$\gamma=10^{-6}$. The mean values of the velocity, calculated over $3000$ 
trajectories, are shown in (c)-(f) and are related to the cases 
$\gamma=0.0,10^{-6}, 10^{-5}, 10^{-4}, 10^{-1}$.} 
\label{velocity1}
\end{figure}

\begin{figure}[htb]
\begin{center}
 \includegraphics*[width=8.0cm,angle=0]{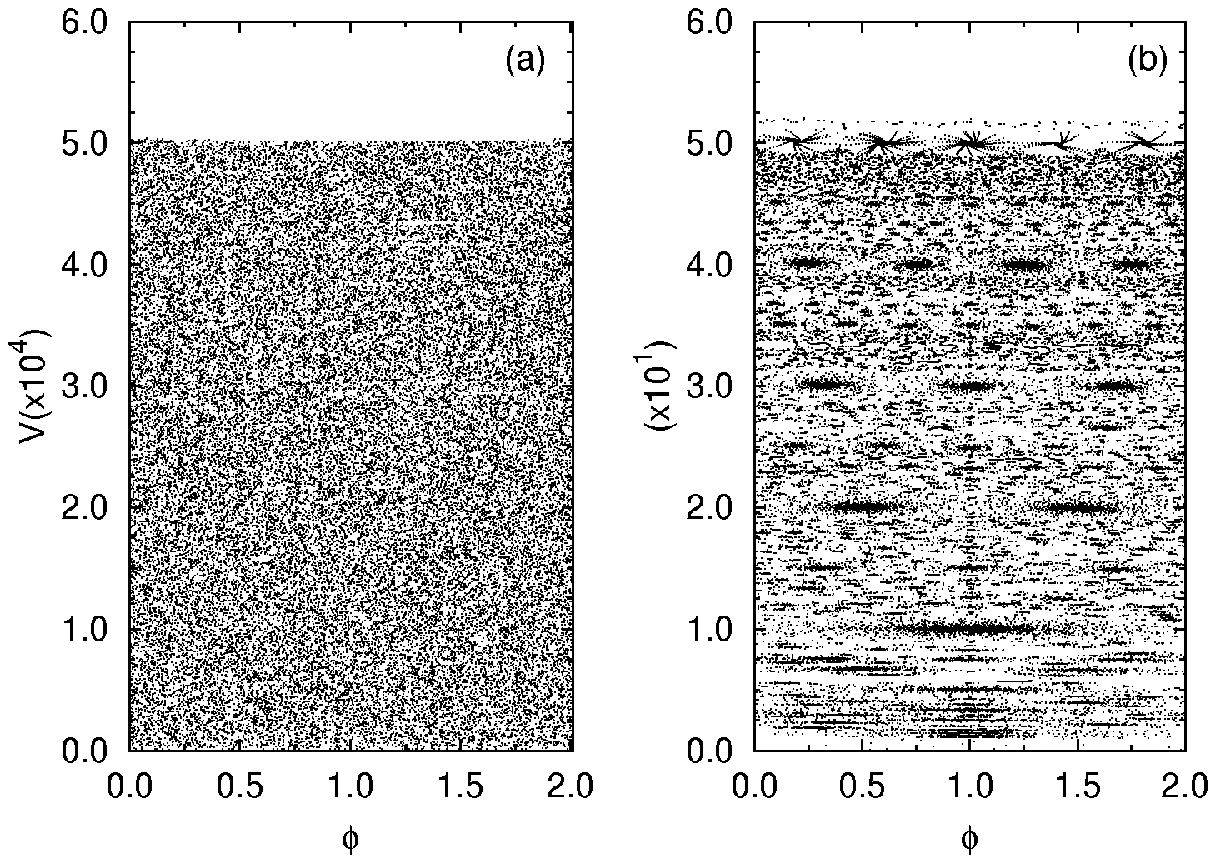}
 \includegraphics*[width=8.0cm,angle=0]{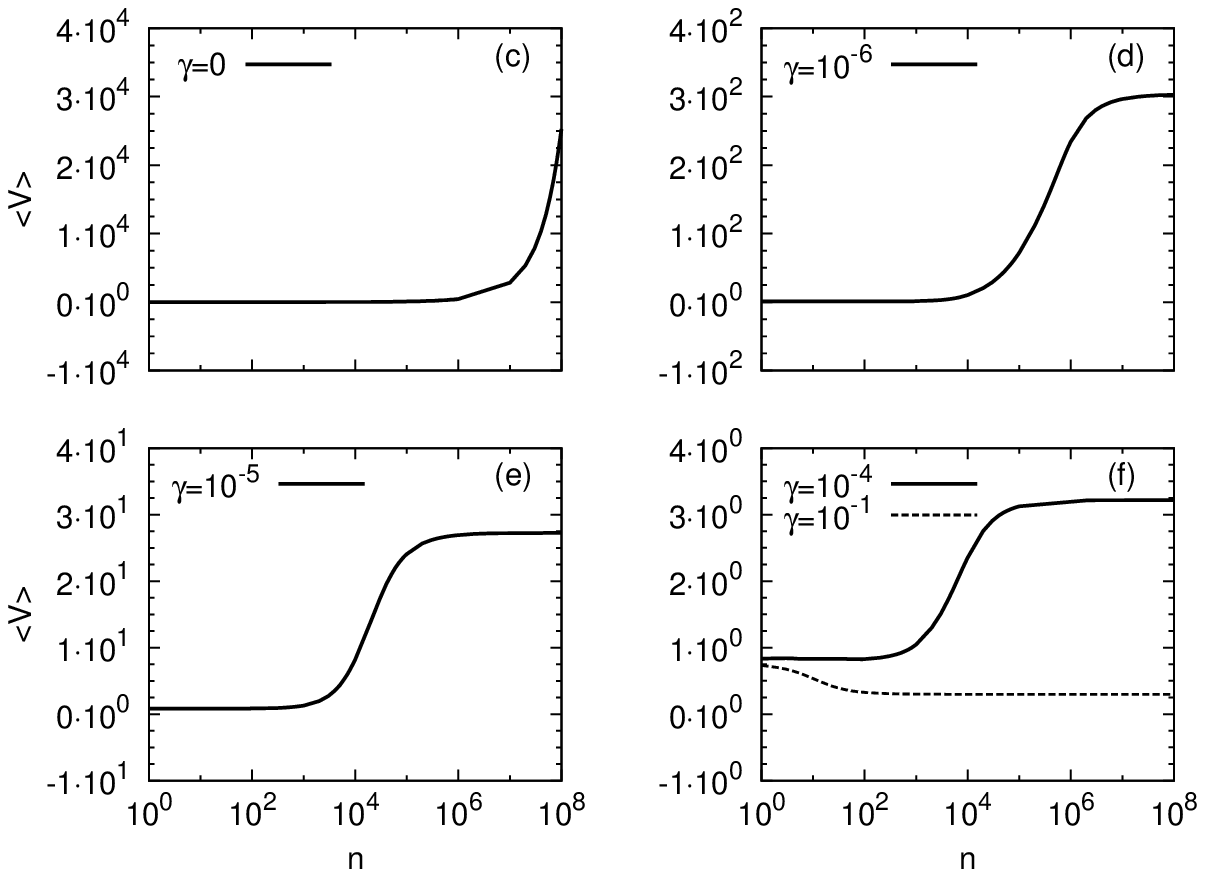}
\end{center}
\caption{\sl Evolution of $500$ chaotic orbits on the phase space
 $V\times\phi$ for the parameters $\mu=10$, $v_0=0.2$ for
case ${\bf H}\, (\Delta\eta=0.01, \eta_1=1.00, \eta_2=1.01$). For 
Fig.~(a) $\gamma=0.0$ and (b) 
$\gamma=10^{-6}$. The mean values of the velocity, calculated over $3000$ 
trajectories are shown in (c)-(f) and are related to the cases 
$\gamma=0.0,10^{-6}, 10^{-5}, 10^{-4}, 10^{-1}$.}
\label{velocity2}
\end{figure}

\begin{figure}[htb]
\begin{center}
 \includegraphics*[width=8.0cm,angle=0]{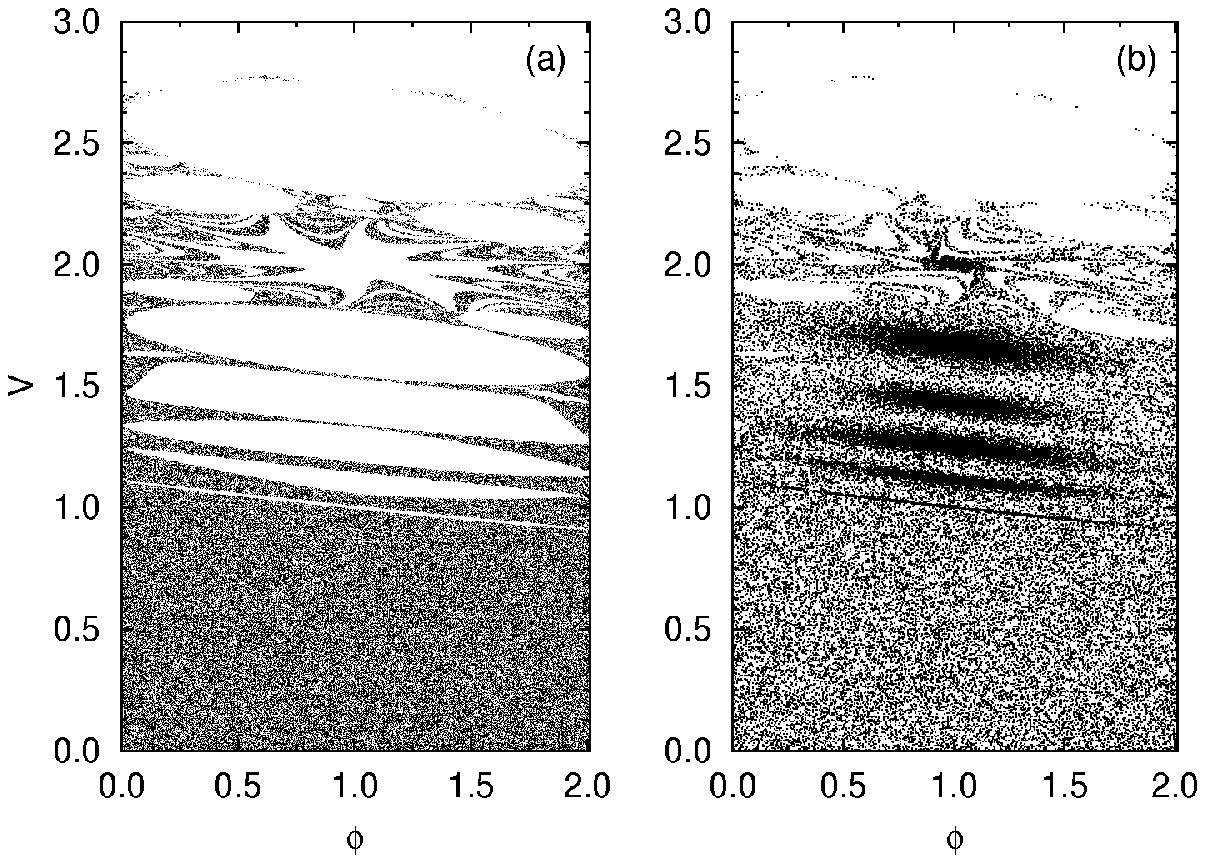}
 \includegraphics*[width=8.0cm,angle=0]{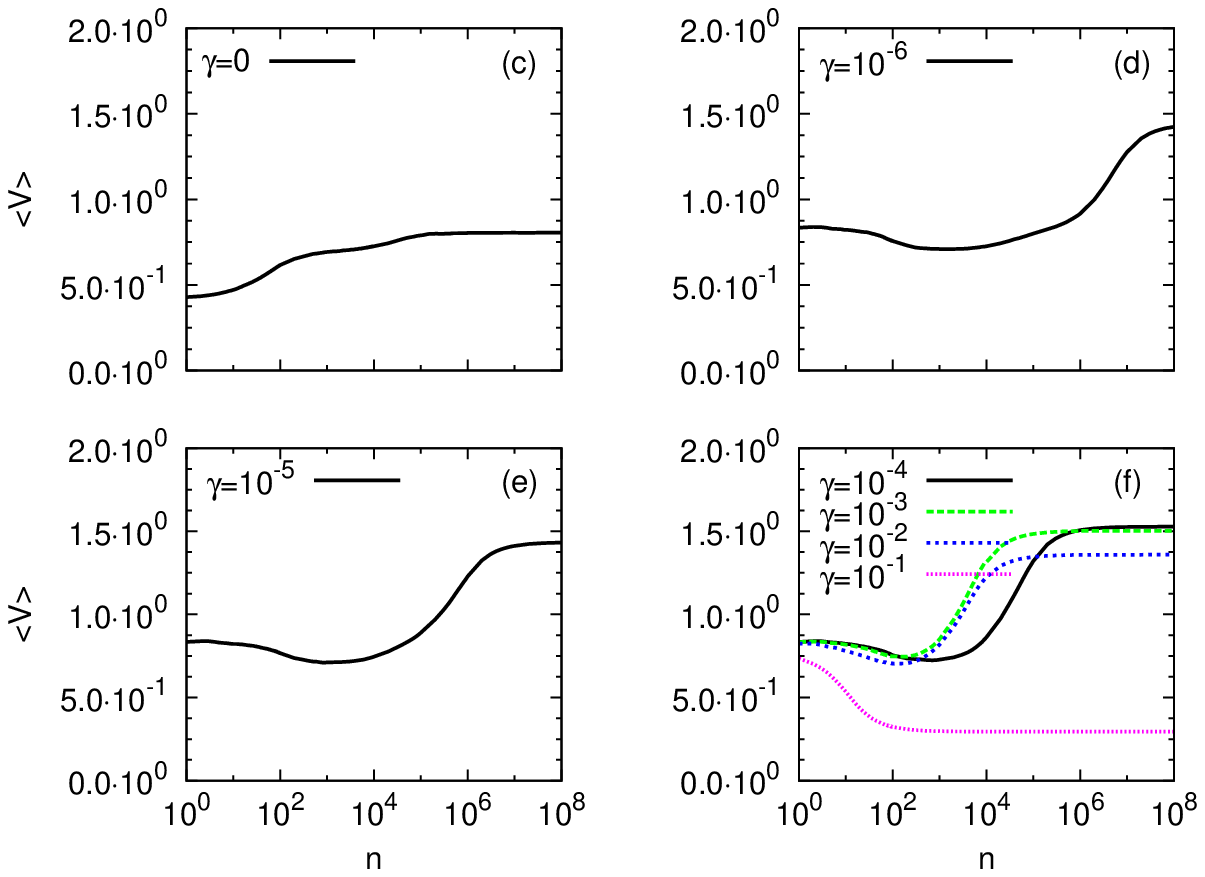}
\end{center}
\caption{\sl Evolution of $500$ chaotic orbits on the phase space
 $V\times\phi$ for the parameters $\mu=10$, $v_0=0.2$ for case
${\bf D}\, (\Delta\eta=-0.01, \eta_1=1.01, \eta_2=1.00$). For Fig.~(a) 
$\gamma=0.0$ and (b) $\gamma=10^{-6}$.
The mean values of the velocity, calculated over $3000$ trajectories, 
are shown in (c)-(f) and are  related to the cases 
$\gamma=0.0,10^{-6}, 10^{-5}, 10^{-4}, 10^{-3}, 10^{-2}, 10^{-1}$.}
\label{velocity3}
\end{figure}

The time asymmetry of the oscillating wall in Eqs.~(\ref{map_1}) and
(\ref{map_2})  is controlled by varying the parameters ($\eta_1,\eta_2$).
The  {\it deformed} sawtooth (Ratchetlike) is obtained when
$\eta_1\ne\eta_2$. Figures \ref{bilhar}(a)-(c)
show the time behaviour of the oscillating wall (the pulse) for different 
values of the asymmetry $\Delta\eta=\eta_2-\eta_1$. The deformed sawtooth
pulse is obtained when $\Delta\eta\ne0.0$. Such pulses can be easily obtained 
from pulse generators. For the case of no dissipation three different motions 
were obtained and controlled \cite{cesarJCP09}. They are ${\bf A:}$ for 
$\Delta\eta=0$ (symmetric sawtooth) the {\it accelerated} case 
[see Fig.~\ref{velocity1}(a)]; ${\bf D:}$ for $\Delta\eta<0$(asymmetric 
sawtooth) the {\it decelerated} case [Fig.~\ref{velocity2}(a)]; and ${\bf H:}$ 
for $\Delta\eta>0$ (asymmetric sawtooth) the {\it hyperaccelerated} case 
[Fig.~\ref{velocity3}(a)]. This classification, different from 
\cite{loskutov07}, is based on how fast the average velocity grows or 
decreases. At next we discuss separately the effects of 
dissipation in each case. To do this we show the corresponding phase space 
dynamics and determine the mean particles velocity at a given time $n$ from 

\begin{equation}
 \langle V \rangle(n) = 
       \frac{1}{n+1}\sum_{i=0}^{n}\frac{1}{\xi}\sum_{j=1}^{\xi}V_{n,j},
\end{equation}
where the index $i$ refers to the $i$th iteration of the sample $j$, and
$\xi$ is the number of initial conditions.  We iterate the 
map~(\ref{map_1}) or (\ref{map_2}) for times $n=1\times 10^8$ and $3000$ 
initial conditions in the interval $0<\phi\le\eta_1+\eta_2$ and 
$0\le V\le10^{-3}$. 

First we consider the accelerated mode which is similar to the simplified 
model studied by Lichtenberg and Lieberman~\cite{lichtenberg72,lichtenberg92}, 
where the harmonic force was considered. This case is shown in 
Fig.~\ref{velocity1}. As the particle velocity increases, regular island are 
observed and $\langle V \rangle$ increases slowly until $\sim 10$ 
[Fig.~\ref{velocity1}(a) and (c)]. The regular islands prevent the particle 
velocity to 
increase very fast. We mention that all initial conditions start inside the 
chaotic region at low velocities. The growth rate of $\langle V \rangle$ depends 
on the number of regular islands inside the phase space. For very small 
dissipations ($\gamma=10^{-6}$) the regular islands are transformed into sinks,
as can be observed by comparing Fig.~\ref{velocity1}(a) and Fig.~\ref{velocity1}(b)
(observe the dark regions), and attract the chaotic trajectories which pass
nearby. As a consequence the mean velocity cannot increase as before (compare 
Fig.~\ref{velocity1}(c) and (d)). This effect increases for higher values 
of the dissipation parameter, as can be observed in Figs.~\ref{velocity1}(e)-(f).
Therefore in case ${\bf A}$ the dissipation decreases the mean velocity 
(to a constant plateau) for any values of $\gamma$ when compared to the 
non-dissipative case.
There are two additional information we can get from Figs.~\ref{velocity1}(d)-(f),
the saturation time ($t_s$), where the mean velocity starts to be constant in
time and the saturated mean velocity ($v_s$). Plotting both quantities as a 
function of $\gamma$ in the interval $10^{-6}\le\gamma\le 10^{-1}$ we obtain 
the decay $t_c\sim \gamma^{-1.0}$ and $v_c\sim \gamma^{-0.3}$.

The second case to discuss is the dissipation in the 
{\it hyperaccelerated} case ${\bf H}$ [see Fig.~\ref{velocity2}(a)]. For a 
very small asymmetry $\Delta\eta=0.01$ the phase space is totally filled and 
no regular islands are observed [Fig.~\ref{velocity2}(a)]. As shown 
in \cite{cesarJCP09}, the Lyapunov exponents approach zero for this case. The 
corresponding $\langle V \rangle$ increases very fast until 
$\langle V \rangle\sim 3\times 10^{4}$ [see Fig.~\ref{velocity2}(c)], showing 
that the accelerated mode is enhanced when compared to the case ${\bf A}$. By 
adding small dissipation ($\gamma=10^{-6}$) into the system we observe in 
Fig.~\ref{velocity2}(b) that again many sinks appear. In this case however, since
no regular islands were observed for $\gamma=0$, we are not able to relate the
sinks to regular islands. On the other hand, we can say that the dissipation is
able to find attracting points which were not visible in the conservative limit.
Also here small dissipation prevents the hyperacceleration of particles. The 
mean particles velocity also decreases (to a constant value) due to the small 
sinks which 
appear in the phase space shown in Fig.~\ref{velocity2}(b). For higher values 
of the dissipation parameter the maximal mean velocity decreases. Observe 
Fig.~\ref{velocity2}(d)-(f) for $\gamma=0.0,10^{-6}, 10^{-5}, 10^{-4}, 10^{-1}$.
For the hyperaccelerated case we found that the saturated quantities decay like
$t_c\sim \gamma^{-1.1}$ and $v_c\sim \gamma^{-0.6}$.

Now we discuss the decelerated case ${\bf D}$ ($\Delta\eta=-0.01$). Here many 
regular islands appear in phase space [see Fig.~\ref{velocity3}(a)] which again 
prevent the acceleration to increase without bounds, as in case ${\bf A}$. The
regular islands are very large so that the maximal instantaneous 
velocity is around $\sim2.5$, while for case ${\bf A}$ it was $\sim 70.0$. 
The corresponding $\langle V \rangle$ remains here almost constant 
[see Fig.~\ref{velocity3}(c)]. As in the other cases, when small 
dissipation is added ($\gamma=10^{-6}$) into the system, regular islands
are transformed into sinks. However here the mean velocity {\it increases}
[see Fig.\ref{velocity3}(d)].
This increasing in the mean velocity is easy to explain. Since the first 
(from below) four islands from the conservative system are very large
(see  Fig.\ref{velocity3}(a)),
almost all initial conditions cannot reach velocities higher than $\sim 1.1$. 
It is a kind of upper partial barrier for the velocities. However, when small 
dissipation is added, these four island are transformed into sinks so that the 
chaotic trajectories can penetrate the islands and the upper barrier does not 
exist anymore.  This makes the mean velocity to increase because the
position of the sinks are higher (in velocity) than the upper barrier. Such 
increase in the mean velocity when dissipation is present was also observed in 
\cite{ricardo}. For the decelerated case we found for the saturated time 
$t_c\sim \gamma^{-1.2}$ while $v_c$ is constant in the interval
$10^{-6}\le\gamma\le10^{-2}$ and decreases for $\gamma=0.1$.

To conclude, for a long time the Fermi acceleration has been studied in 
different models and applications
\cite{fermi49,ulam61,lichtenberg72,lichtenberg92,leonel04,leonel04-prl,
leonel05,egydio06,karlis06,karlis07,pustylnikov77,steane95,schleich98,
saif05,milovanov01,michalek99,veltri04,kobayakawa02,malkov98,cole06},
but just recently \cite{cesarJCP09} a deformed sawtooth (Ratchetlike) pulse
was proposed to control the Fermi de(acceleration).  Changing the asymmetry 
parameter from the Ratchetlike pulse it was possible to get Fermi 
hyperacceleration and deceleration. By switching the pulse between 
hyperacceleration and deceleration modes an accurate control of the particles 
velocity was achieved. Beside the remarkable control of velocities obtained 
in the dissipation free problem, we observe here that dissipation effects
transform regular islands into sinks, change the mean velocity and reduce the 
ability to control accurately the particle dynamics. The saturation time decays 
with a power law  $t_c\sim\gamma^{-\beta}$, where $\beta$ is close to $1.0$ 
($\pm 0.01$) for all considered cases. On the other hand, the saturated mean 
velocity also obeys the power law decay, but with exponents $\beta\sim0.6$ for 
the case ${\bf H}$, and  $\beta\sim0.3$ for the case ${\bf A}$. In the 
decelerated case the saturated velocity is almost constant for small 
dissipations.  
Our results motivate further analysis related to the control of Fermi
de(acceleration) for larger Ratchetlike asymmetries of the ac driven pulse in 
the presence of dissipation \cite{cesar-beims09} and in the hopping wall 
approximation. In the later case we expect to
obtain a better efficiency to control the particles velocity since the 
the energy gain per collision is not underestimated as in the present model. 
In addition, it would be interesting to analyse the implementation of the
Ratchetlike ac driven pulse applied to the walls of quasi 
one-dimensional billiards which are coupled to thermal baths at different
temperatures. In such cases it is desirable to optimize and achieve the 
control of directed heat conduction \cite{klages07}, which could leave to an 
important increase of the thermal efficiency in physical devices such as
rectifiers and thermal transistors \cite{casati08}.

\section{Acknowledgments}
The authors thank CNPq and FINEP (under project CTINFRA-1) for financial
support.

\end{document}